\documentclass[onecolumn,aps,prd,preprintnumbers,showpacs,
superscriptaddress,nofootinbib,amsmath,amssymb,floats,floatfix,
showkeys,notitlepage,longbibliography]{revtex4-1}

\usepackage{graphicx}
\usepackage{subfigure}
\usepackage{orcidlink}
\usepackage{hyperref}
\hypersetup{colorlinks=true,linkcolor=blue,urlcolor=blue,citecolor=blue}
\usepackage{bm}
\usepackage{bbm}

\begin{document}

\title{Quasinormal Mode Spectroscopy via Horizon-Brightened Quantum Optics}

\author{Ali \"Ovg\"un \orcidlink{0000-0002-9889-342X}}
\email{ali.ovgun@emu.edu.tr}
\affiliation{Physics Department, Faculty of Arts and Sciences, Eastern Mediterranean University, Famagusta, 99628 North
Cyprus via Mersin 10, Turkiye.}
\date{\today }

\begin{abstract}
We develop a quantum optical framework for probing black hole quasinormal modes (QNMs) using two-level atoms in the spirit of the horizon-brightened acceleration radiation (HBAR) program. Starting from the QNM contribution to the Wightman function of a scalar field on a static, spherically symmetric black hole background, we derive the response function of a two-level Unruh--DeWitt detector following simple trajectories (static at fixed radius, with comments on radial free fall). The QNM sector imprints a set of Lorentzian resonances in the detector spectrum at the redshifted real parts of the QNM frequencies, with widths determined by the imaginary parts. We then treat a single dominant QNM as an effective non-Hermitian cavity mode coupled to an ensemble of driven two-level atoms, and derive a master equation of Dicke laser type. The resulting lasing threshold condition depends explicitly on the QNM damping rate, providing a direct quantum optical interpretation of the imaginary part of the QNM frequency. Specializing to the Schwarzschild geometry, we express the resonant frequencies, linewidths, and threshold in terms of photon-sphere data in the eikonal limit. We discuss several extensions and propose our framework as a unifying language connecting black hole ringdown, near-horizon conformal quantum mechanics, and quantum optics, thereby enriching the emerging program of black hole spectroscopy in the gravitational-wave era.
\end{abstract}

\maketitle

\section{Introduction}

Quantum field theory in curved spacetime predicts that horizons endow the vacuum with a highly nontrivial structure, giving rise to phenomena such as Unruh and Hawking radiation~\cite{Unruh:1976db,Birrell:1982ix,Crispino:2007eb}. An especially sharp way to make this structure operational is through the response of localized particle detectors, typically modeled as Unruh--DeWitt (UDW) two-level systems coupled to quantum fields~\cite{DeWitt:1979ud,Takagi:1986kn,Louko:2008vy}. In this language, the notion of particles and temperature is encoded in detector transition probabilities and open-system dynamics rather than in global field modes, and the effects of acceleration, curvature, and horizons are directly reflected in the detector response.

Over the last decade, this operational viewpoint has merged with the toolbox of quantum optics and relativistic quantum information. Atoms, cavities, and interferometers are now widely used as theoretical probes of near-horizon physics and accelerated motion. A paradigmatic development is the ``horizon-brightened acceleration radiation'' (HBAR) program, in which clouds of two-level atoms falling into a Schwarzschild black hole in the Boulware vacuum are shown to emit a spectrum that is effectively thermal at the Hawking temperature, despite the global state not being thermal~\cite{ScullyHBAR1,Camblong:2020pme,Azizi:2020gff,Azizi:2021qcu,Azizi:2021yto,Scully:2022bun,Scully:2022pov,Das:2023rwg,Ordonez:2025sqp}. In this quantum-optical picture, the black hole acts as a kind of broadband quantum amplifier: infalling or accelerated atoms play the role of gain media, and their excitation and emission probabilities encode detailed information about the spacetime geometry and its thermodynamic properties. Closely related ideas have been sharpened in the near-horizon conformal quantum mechanics (CQM) framework, where the radial dynamics effectively reduces to an inverse-square potential in \(0+1\) dimensions~\cite{Camblong:2005my,Camblong:2022jbg,Eissa:2025vhs}.

This emerging ``relativistic quantum optics’’ program sits at the intersection of quantum information, quantum field theory, and gravitation. A broad class of works has analyzed accelerated or freely falling atoms, UDW detectors, and mirrors in flat and curved backgrounds, clarifying the relation between acceleration, detector response, and the equivalence principle~\cite{Ben-Benjamin:2019opz,Fulling:2018lez,Tjoa:2018xre,Chakraborty:2019ltu,Lopp:2018lxl,Lopp:2021zvo,Svidzinsky:2018jkp,Svidzinsky:2019jqr,Sadurni:2021rwy,Good:2021ffo,Das:2022qcr,Prokhorov:2019sss,Das:2022qpx}. These studies have been extended to Casimir setups and freely falling cavities or plates in gravitational fields~\cite{Sorge:2018zfd,Costa:2020asn,Masood:2024glj}, to light propagation in effective curved spaces and optical analogs~\cite{Xu:2021cbw,Pan:2024jix,Rozenman:2024ymh}, and to gravitational-wave and dynamical backgrounds where light–matter coupling is treated within Jaynes–Cummings–type models~\cite{Sorge:2023tyn,McMaken:2024vsn,MasoodASBukhari:2023flr,Wang:2025jvb}. Collectively, these works demonstrate that relativistic motion and gravity imprint measurable signatures on atomic transition rates, interference fringes, and quantum correlations.

A central theme in this literature is the status of the equivalence principle and the gravitational mass of composite quantum systems. By comparing accelerated and freely falling detectors, mirror–atom configurations, and cavity setups, one can ask to what extent acceleration radiation and related phenomena are purely kinematic, or whether genuinely gravitational imprints survive~\cite{Fulling:2018lez,Tjoa:2018xre,Zych:2018bmk,Chatterjee:2021fue,Chatterjee:2021fsg,Chakraborty:2019ltu,Barman:2021oum,Liu:2019soq}. Near-horizon analyses have revealed universal instability and redshift mechanisms underlying quantum thermality~\cite{Dalui:2020qpt,Sen:2022cdx,Dalui:2023lit,Good:2021ffo}, while nonlocal and generalized field theories highlight how the Unruh–Fulling effect is modified by ultraviolet completions~\cite{Das:2022qcr,Das:2022qpx}. In parallel, density-operator methods have been developed to treat acceleration and vorticity within a unified thermodynamic formalism~\cite{Prokhorov:2019sss}.

From a quantum-information perspective, acceleration and horizon physics have been explored through entanglement, steering, and Bell tests with relativistic detectors. Entanglement harvesting by pairs of accelerated or freely falling atoms, including along null trajectories and in black-hole backgrounds, reveals how curvature and horizon structure affect nonlocal correlations~\cite{Gallock-Yoshimura:2021yok,Barman:2021kwg,Barman:2023rhd,Barman:2021oum,Stargen:2025siq}. Quantum steering ellipsoids and entropic measures characterize how Unruh noise reshapes quantum resources~\cite{Maleki:2021idq,You:2018qgr,Tian:2023sfz}, while Bell-inequality violations in dynamical Casimir and circuit-QED platforms offer promising routes to simulate horizon-induced correlations in the laboratory~\cite{Chatterjee:2022pkj,Svidzinsky:2018jkp,Svidzinsky:2019jqr,Chatterjee:2021fsg}. Proposals such as Ramsey interferometry witnesses of acceleration radiation~\cite{Costa:2020asn,Lopes:2021ajs}, shock-wave–induced quantum memories~\cite{Majhi:2021byc}, and quantum amplifiers powered by black holes or horizons~\cite{Misra:2023lxt,Allen:2019muo,Scully:2022bun,Longhi:2025ljw} further deepen the analogy between gravitational horizons and nontrivial optical media.

Within this broader context, the HBAR program refines the UDW picture by focusing on atoms falling into, or hovering near, black-hole horizons and related causal boundaries. The emission and absorption spectra, together with the associated entropy production, encode a ``horizon brightening’’ of acceleration radiation that depends on the near-horizon conformal structure and on the microscopic details of the coupling~\cite{ScullyHBAR1,ScullyHBAR2,Camblong:2020pme,Azizi:2020gff,Sen:2022tru,Sen:2022cdx,Das:2023rwg,Azizi:2021qcu,Azizi:2021yto,Eissa:2025vhs}. This framework has now been applied to a wide variety of geometries, including quantum-corrected and braneworld black holes~\cite{Sen:2022tru,Das:2023rwg,Jana:2024fhx,Jana:2025hfl}, Kerr–Newman and rotating backgrounds~\cite{Azizi:2020gff,Sen:2023zfq,McMaken:2024vsn}, and scenarios where dark energy or dark matter fields modify the near-horizon region~\cite{Bukhari:2022wyx,Bukhari:2023yuy,Barman:2023rhd,MasoodASBukhari:2023flr}. Derivative couplings and their impact on acceleration radiation and HBAR entropy have been explored in detail~\cite{Das:2025rzz,Pantig:2025okn}, and corrections arising from generalized uncertainty principles and renormalization-group improvements of the metric have been shown to leave characteristic imprints on HBAR entropy~\cite{Sen:2022tru,Jana:2024fhx,Jana:2025hfl,Ovgun:2025isv}.

Generalizations to Lorentz-violating and higher-curvature spacetimes open an additional window on fundamental physics. Acceleration radiation and HBAR-type quantities have been investigated for atoms in Lorentz-violating black-hole backgrounds, including Kalb–Ramond geometries, in order to identify potential observational signatures of symmetry breaking in the quantum regime~\cite{Tang:2025eew,Rahaman:2025grm,Rahaman:2025mrr}. Casimir-like probes and optical analogs in modified gravity provide complementary constraints on higher-derivative corrections and effective couplings~\cite{Masood:2024glj,Xu:2021cbw,Pan:2024jix}. At the same time, refined phase-space techniques and Wigner distributions in Rindler spacetime clarify the role of nonvacuum states and quantum coherence in acceleration phenomena~\cite{Ullinger:2022xmv,Dubey:2024abz,Wang:2025jvb}. Experimental and numerical advances, from water-wave analog horizons and accelerated-electron proposals for Unruh radiation to machine-learning–enhanced signal extraction in neutrino experiments, suggest that subtle acceleration-induced quantum effects may be within reach~\cite{Gregori:2023tun,Rozenman:2024ymh,Maksimovic:2023ttj}.

In parallel, the detection of gravitational waves from binary black-hole mergers by LIGO and Virgo has inaugurated the era of black-hole spectroscopy, in which the ringdown phase of the signal is analyzed to extract the quasinormal-mode (QNM) spectrum and test general relativity in the strong-field regime~\cite{Berti:2009kk,Konoplya:2011qq}. QNMs encode the characteristic relaxation of a perturbed black hole through complex frequencies
\begin{equation}
  \omega_{n\ell} = \Omega_{n\ell} - i\,\Gamma_{n\ell},
\end{equation}
whose real and imaginary parts determine, respectively, the oscillation frequencies and damping times~\cite{Kokkotas:1999bd,Nollert:1999ji}. The prospect of \emph{black-hole spectroscopy}---inferring the underlying geometry, mass, and spin from these spectral lines---has motivated extensive work on the structure and phenomenology of QNMs.

At the same time, there exists a mature and rapidly evolving body of research on the \emph{classical} optical appearance of black holes, in particular their shadows and strong-field lensing properties. The shape and size of the black-hole shadow, as well as its deformation by spin, deviations from Kerr, higher-dimensional corrections, or braneworld effects, provide an independent probe of the geometry and of possible extensions of general relativity~\cite{Claudel:2000yi,Virbhadra:2002ju,Adler:2022qtb,Vagnozzi:2022moj,Allahyari:2019jqz,Khodadi:2020jij,Atamurotov:2013sca,Atamurotov:2015nra,Abdujabbarov:2015rqa,Abdujabbarov:2017pfw}. Studies of rotating non-Kerr spacetimes, plasmas around compact objects, and Gauss–Bonnet or braneworld black holes show how shadow and lensing observables respond to departures from the standard Kerr paradigm~\cite{Atamurotov:2013sca,Atamurotov:2015nra,Abdujabbarov:2015rqa,Abdujabbarov:2017pfw}. These imaging signatures complement the HBAR and acceleration-radiation program: while shadows and lensing probe the null-geodesic structure and effective refractive properties of the spacetime, quantum-optical probes via infalling atoms and cavities are sensitive to the detailed structure of quantum fields near the horizon. A consistent theoretical framework should ultimately relate these classical and quantum observables, providing a multi-channel spectroscopy of black holes.

These developments strongly suggest that QNMs, Hawking/Unruh physics, and quantum optics are deeply intertwined. On the one hand, QNMs appear as poles of the Green's functions of perturbations in a black-hole background~\cite{Leaver:1986gd,Andersson:1996cm,Casals:2013mpa}. Quasinormal modes (QNMs) have been extensively studied as precise probes of black-hole spacetimes in a wide variety of backgrounds, including Schwarzschild–(A)dS, de Sitter, regular, and modified-gravity geometries, where scalar and gravitational perturbations reveal the impact of topology, nonlinearity, and higher-curvature terms on the spectrum~\cite{Cardoso:2001bb,Zhidenko:2003wq,Aros:2002te,Fernando:2012yw,Toshmatov:2015wga,Rincon:2018sgd,Konoplya:2020bxa,Pantig:2022gih,Fu:2022cul,Fu:2023drp}. The role of field content and rotation has been clarified through detailed analyses of massive scalar and vector fields, as well as eikonal-limit studies in Kerr and slowly rotating backgrounds, which connect QNMs to superradiant instabilities and unstable null orbits~\cite{Dolan:2007mj,Dolan:2010wr,Pani:2012bp,Frolov:2018ezx,Konoplya:2017wot}. On the methodological side, higher-order WKB schemes, continued-fraction techniques, and autocorrelation-based approaches provide efficient and accurate tools for computing spectra and greybody factors, and for assessing the physical significance of individual modes~\cite{Konoplya:2019hlu,Daghigh:2022uws,Cardoso:2001bb,Qian:2021aju,Daghigh:2020jyk}. More recently, loop-quantum-gravity–motivated corrections, regular black-hole models, and topological geometries have been used as test beds to investigate stability, spectral (in)stability, and potential echo signatures in the QNM spectrum~\cite{Daghigh:2020fmw,Daghigh:2020mog,Aros:2002te,Dias:2015wqa,Shen:2025yiy}.
On the other hand, detector-based formulations of quantum field theory probe precisely those Green's functions through response functions and open-system dynamics. Yet a systematic quantum-optical description in which QNMs play the role of effective cavity modes interacting with atomic degrees of freedom has not been fully developed.

In this work we propose and analyze such a framework, which we refer to as \emph{HBAR--QNM spectroscopy}. The central idea is to regard the QNM contribution to the Wightman function as an effective discrete sector describing damped bosonic modes, and to couple this sector to a cloud of two-level atoms in the spirit of standard cavity quantum optics~\cite{ScullyZubairy}. This enables us to import familiar concepts such as Lorentzian resonances, lasing thresholds, and Maxwell–Bloch equations into black-hole physics, and to reinterpret the imaginary parts of QNM frequencies as effective loss rates in a quantum-optical setting. More concretely, our main objectives are: To decompose the Wightman function into a QNM sector plus a continuum, where the QNM sector furnishes an effective discrete set of damped ``cavity modes’’ with complex frequencies \( \omega_n = \Omega_n - i\Gamma_n \), whose residues encode their spatial profiles~\cite{Leaver:1986gd,Andersson:1996cm,Casals:2013mpa}. To compute the response of two-level UDW detectors to this QNM sector. For static detectors, we show that QNMs produce Lorentzian resonances in the excitation rate at the locally redshifted QNM frequencies, with widths set by the redshifted damping rates. To treat a dominant QNM as a non-Hermitian bosonic mode coupled to an ensemble of pumped two-level atoms. A standard quantum-optical derivation then yields a Maxwell–Bloch system and a lasing threshold condition in which the imaginary part of the QNM plays the role of cavity loss. To specialize to the Schwarzschild geometry and express the spectroscopic signals in terms of photon-sphere quantities in the eikonal limit~\cite{Cardoso:2008bp}, thereby providing a geometric interpretation of the HBAR–QNM fingerprint.

Our treatment is deliberately idealized and should be viewed as a theoretical laboratory rather than an immediate observational proposal: we do not attempt to model realistic astrophysical environments or backreaction. Instead, our goal is to establish a coherent formal framework in which QNMs, detector response, and quantum-optical concepts can be discussed on equal footing. Within this framework, we can make explicit statements about how QNM damping rates enter lasing thresholds, how the near-horizon thermal behavior encoded in CQM combines with photon-sphere dynamics, and how different black-hole geometries lead to distinct HBAR–QNM spectral fingerprints.

The structure of the paper is as follows. In Sec.~\ref{sec:WightmanQNM} we review the mode decomposition of a scalar field on a static, spherically symmetric black hole and isolate the QNM contribution to the positive-frequency Wightman function. In Sec.~\ref{sec:detector} we compute the response of a two-level UDW detector to this QNM sector and identify the resulting Lorentzian resonances. In Sec.~\ref{sec:lasing} we introduce an effective single-QNM master equation for a cloud of atoms coupled to a damped mode, derive the corresponding Maxwell–Bloch equations, and obtain a lasing threshold condition explicitly involving the QNM damping rate. In Sec.~\ref{sec:Schwarzschild} we specialize to the Schwarzschild case, use photon-sphere data to approximate the QNM spectrum, and discuss the resulting HBAR–QNM spectral phenomenology. Section~\ref{sec:figures} outlines illustrative figures and qualitative interpretations of our results, and we summarize and discuss extensions in Sec.~\ref{sec:conclusions}.

Throughout, we work in units \(c = G = k_B = \hbar = 1\), unless otherwise stated, and adopt the mostly-plus metric signature \((-,+,+,+)\).

\section{Quasinormal-mode contribution to the Wightman function}
\label{sec:WightmanQNM}

In this section we briefly review the mode decomposition of a minimally coupled scalar field in a static, spherically symmetric black hole spacetime and recall how QNMs arise as poles of the Green's function. Our goal is to isolate a QNM contribution to the positive-frequency Wightman function that will later serve as an effective discrete sector in the detector response.

We consider a static, spherically symmetric black hole background of the form
\begin{equation}
  ds^2 = - f(r)\,dt^2 + f(r)^{-1}\,dr^2 + r^2\,d\Omega^2,
  \label{eq:metric}
\end{equation}
where \( d\Omega^2 \) is the metric on the unit 2-sphere. The horizon is located at \( r = r_h \), where \( f(r_h) = 0 \), and the corresponding surface gravity is
\begin{equation}
  \kappa = \frac{1}{2} f'(r_h).
\end{equation}
For concreteness we will later specialize to the Schwarzschild case, \( f(r) = 1 - 2M/r \), but for now we keep \( f(r) \) general.

We consider a real, minimally coupled scalar field \( \Phi \) obeying the Klein--Gordon equation
\begin{equation}
  \Box \Phi = \frac{1}{\sqrt{-g}}\partial_\mu\bigl(\sqrt{-g}\,g^{\mu\nu}\partial_\nu \Phi\bigr) = 0,
  \label{eq:KG}
\end{equation}
and expand it in a complete set of normal modes adapted to the spherical symmetry and stationarity of the background.


We first decompose the field in spherical harmonics,
\begin{equation}
  \Phi(t,r,\Omega)
  = \sum_{\ell=0}^{\infty}\sum_{m=-\ell}^{\ell}
    \frac{1}{r}\,\phi_{\ell m}(t,r)\,Y_{\ell m}(\Omega),
  \label{eq:spherical-decomp}
\end{equation}
where \( \Omega \) denotes the angular coordinates on the sphere and \( Y_{\ell m}(\Omega) \) are the usual spherical harmonics. The prefactor \(1/r\) is chosen so that the radial modes satisfy a Schr\"odinger-type equation.

Next we introduce the tortoise coordinate \( r_* \), defined by
\begin{equation}
  \frac{dr_*}{dr} = \frac{1}{f(r)}.
  \label{eq:tortoise}
\end{equation}
This maps the exterior region \( r \in (r_h,\infty) \) to \( r_* \in (-\infty,\infty) \), with the horizon at \( r_* \to -\infty \) and spatial infinity at \( r_* \to +\infty \). In terms of \(r_*\), the radial dynamics takes a form reminiscent of scattering in one dimension.

Substituting Eq.~\eqref{eq:spherical-decomp} into the Klein--Gordon equation~\eqref{eq:KG}, we find that each partial wave \( \phi_{\ell m}(t,r) \) satisfies a \(1+1\)-dimensional wave equation,
\begin{equation}
  \left( \frac{\partial^2}{\partial t^2}
       - \frac{\partial^2}{\partial r_*^2}
       + V_\ell(r) \right)
  \phi_{\ell m}(t,r) = 0,
  \label{eq:radialwave}
\end{equation}
where the effective potential for a minimally coupled scalar field is
\begin{equation}
  V_\ell(r) = f(r)\left[\frac{\ell(\ell+1)}{r^2}
                      + \frac{f'(r)}{r}\right].
  \label{eq:scalar-potential}
\end{equation}
The first term is the centrifugal barrier, while the second encodes curvature effects. For other spins, the potential is modified in a known way~\cite{Regge:1957td,Zerilli:1970se,Konoplya:2011qq}, but the overall structure of Eq.~\eqref{eq:radialwave} remains intact: each multipole propagates in an effective potential barrier that typically peaks near the photon sphere and decays towards the horizon and infinity.

It is often convenient to work in the frequency domain. Writing
\begin{equation}
  \phi_{\ell m}(t,r) = \int_{-\infty}^{+\infty}\frac{d\omega}{2\pi}\,
   e^{-i\omega t}\,\psi_{\ell\omega}(r),
\end{equation}
we obtain the ordinary differential equation
\begin{equation}
  \left(-\frac{d^2}{dr_*^2} + V_\ell(r)\right)\psi_{\ell\omega}(r)
  = \omega^2\,\psi_{\ell\omega}(r).
  \label{eq:radial-ODE}
\end{equation}
This equation defines a scattering problem in the potential \( V_\ell(r) \), with incoming and outgoing components at the horizon and infinity.


The retarded Green's function for a fixed multipole \(\ell\) can be represented in the time and frequency domains as
\begin{equation}
  G_\ell^{\rm ret}(t,r;r')
   = \int_{-\infty}^{+\infty}\frac{d\omega}{2\pi}\,
     e^{-i\omega t}\,
     g_\ell(\omega; r_*, r_*'),
  \label{eq:retarded-G}
\end{equation}
where \( g_\ell(\omega; r_*, r_*') \) is built from two independent solutions of Eq.~\eqref{eq:radial-ODE} satisfying appropriate boundary conditions (purely ingoing at the horizon and purely outgoing at infinity) and stitched together via the Wronskian~\cite{Leaver:1986gd,Andersson:1996cm}.

Quasinormal modes are defined as solutions of the homogeneous equation,
\begin{equation}
  \left(-\frac{d^2}{dr_*^2} + V_\ell(r)\right) u_{n\ell}(r_*)
  = \omega_{n\ell}^2\,u_{n\ell}(r_*),
  \label{eq:QNM-eq}
\end{equation}
subject to the boundary conditions of purely ingoing waves at the horizon and purely outgoing waves at spatial infinity. These conditions select a discrete set of complex frequencies
\begin{equation}
  \omega_{n\ell} = \Omega_{n\ell} - i\,\Gamma_{n\ell},
  \qquad \Gamma_{n\ell}>0,
  \label{eq:QNM-frequencies}
\end{equation}
reflecting the fact that perturbations decay over time due to radiation escaping to infinity or being absorbed by the horizon~\cite{Berti:2009kk,Konoplya:2011qq}.

Mathematically, QNMs correspond to poles of the Green's function in the complex frequency plane. One can deform the contour of integration in Eq.~\eqref{eq:retarded-G} into the lower half-plane, picking up contributions from these poles and from branch cuts associated with the continuous spectrum~\cite{Leaver:1986gd,Andersson:1996cm,Casals:2013mpa}. This yields a decomposition of the Green's function into a QNM sum plus a continuum (or ``tail'') contribution. The same structure carries over to the positive-frequency Wightman function
\begin{equation}
  G^+(x,x') = \langle 0|\Phi(x)\Phi(x')|0\rangle,
\end{equation}
where \(|0\rangle\) denotes the chosen vacuum state (Boulware, Unruh, Hartle--Hawking, etc.).

Schematically, we may write
\begin{equation}
  G^+(x,x') = G^+_{\rm QNM}(x,x') + G^+_{\rm cont}(x,x'),
  \label{eq:Gplus-decomp}
\end{equation}
where \(G^+_{\rm QNM}\) is a discrete sum over QNM poles and \(G^+_{\rm cont}\) is a continuum contribution sensitive to the choice of quantum state. The QNM part captures the dominant late-time ringdown behavior, while the continuum encodes power-law tails and thermal features.

A more explicit expression for the QNM contribution can be written as
\begin{equation}
  G^+_{\rm QNM}(x,x') \simeq
    \sum_{\ell m}\sum_{n}
    \mathcal{N}_{n\ell}\,
    \frac{u_{n\ell}(r_*)u_{n\ell}(r_*')}{r r'}\,
    Y_{\ell m}(\Omega)Y_{\ell m}^*(\Omega')\,
    e^{-i\omega_{n\ell}(t-t')},
  \label{eq:GQNMfull}
\end{equation}
where the coefficients \( \mathcal{N}_{n\ell} \) are determined by the residues of the Green's function at the QNM poles and encode an appropriate normalization of the modes~\cite{Leaver:1986gd,Andersson:1995zk,Andersson:1996cm}. Equation~\eqref{eq:GQNMfull} provides a QNM-based approximation to the two-point function that is valid in regimes where the discrete ringing dominates over the continuum.


For our purposes it is useful to focus on the correlation function evaluated at a fixed spatial point, which will later be taken as the location of a detector. We choose a fixed radius \( r_0 \) and angular position \( \Omega_0 \), and define the time difference
\begin{equation}
  \Delta t := t - t'.
\end{equation}
Evaluating Eq.~\eqref{eq:GQNMfull} at \( (r,\Omega)=(r_0,\Omega_0) \) and \( (r',\Omega')=(r_0,\Omega_0) \), and summing over \(m\), we can absorb the angular factors and mode residues into effective coefficients
\begin{equation}
  B_n(r_0) := \sum_{\ell}
     \mathcal{N}_{n\ell}\,
     \frac{|u_{n\ell}(r_*^0)|^2}{r_0^2}\,
     \sum_{m=-\ell}^{\ell}|Y_{\ell m}(\Omega_0)|^2,
\end{equation}
where \( r_*^0 = r_*(r_0) \). We have reindexed the modes so that the label \(n\) runs over all relevant QNMs (including the multipole index \(\ell\), if desired). In terms of these coefficients, the QNM contribution to the Wightman function along a fixed spatial point reduces to
\begin{equation}
  G^+_{\rm QNM}(\Delta t; r_0)
  \equiv G^+_{\rm QNM}(t,r_0,\Omega_0; t',r_0,\Omega_0)
  \simeq \sum_{n} B_n(r_0)\,e^{-i\omega_n \Delta t},
  \label{eq:GQNMsimple}
\end{equation}
where we denote the (relabelled) QNM frequencies by
\begin{equation}
  \omega_n = \Omega_n - i\,\Gamma_n,
  \qquad \Gamma_n>0.
\end{equation}
This simple exponential series will be the starting point for our analysis of detector response.

\section{Detector response and QNM resonances}
\label{sec:detector}

We now turn to the response of a two-level Unruh--DeWitt detector coupled to the field along a given trajectory \( x(\tau) \), parameterized by the detector's proper time \( \tau \). Our goal is to show how the QNM sector of the Wightman function, Eq.~\eqref{eq:GQNMsimple}, imprints Lorentzian resonances in the detector's excitation spectrum, localized at redshifted QNM frequencies and with linewidths determined by the damping rates.


A UDW detector is modeled as a two-level quantum system with ground and excited states \( |g\rangle \) and \( |e\rangle \), respectively, separated by an energy gap \( \nu > 0 \). In the interaction picture, the monopole operator of the detector can be written as
\begin{equation}
  m(\tau) = \sigma_- e^{-i\nu\tau} + \sigma_+ e^{i\nu\tau},
\end{equation}
where \( \sigma_+ = |e\rangle\langle g| \) and \( \sigma_- = |g\rangle\langle e| \) are raising and lowering operators on the detector Hilbert space.

The detector couples to the scalar field along its worldline \( x(\tau) \) via the interaction Hamiltonian
\begin{equation}
  H_I(\tau) = g\,\chi(\tau)\,m(\tau)\,\Phi[x(\tau)],
\end{equation}
where \( g \) is a small coupling constant and \( \chi(\tau) \) is a real-valued switching function that controls the duration and smoothness of the interaction~\cite{Louko:2008vy}.

Assuming that the detector is initially in the ground state and the field in a vacuum state \(|0\rangle\), the probability of excitation at leading order in perturbation theory is given by
\begin{equation}
  P_{g\to e}(\nu) = g^2\,\mathcal{F}(\nu),
\end{equation}
where the \emph{response function} is defined as~\cite{Unruh:1976db,Birrell:1982ix,Crispino:2007eb}
\begin{equation}
  \mathcal{F}(\nu) = 
   \int_{-\infty}^{+\infty} d\tau
   \int_{-\infty}^{+\infty} d\tau'\,
   \chi(\tau)\chi(\tau')\,
   e^{-i\nu(\tau-\tau')}\,
   G^+\bigl(x(\tau),x(\tau')\bigr).
  \label{eq:Fgeneral}
\end{equation}
This expression explicitly shows how the detector response is determined by the two-point function of the field evaluated along the detector's trajectory. The exponential factor \( e^{-i\nu(\tau-\tau')} \) implements energy conservation between the detector and the field, while the switching function ensures that the integral is finite and physically well-defined.

In stationary situations, where the Wightman function depends only on the proper-time difference \( \tau-\tau' \) and the switching is slow, it is often convenient to define a transition rate,
\begin{equation}
  \dot{\mathcal{F}}(\nu) = \lim_{T\to\infty}\frac{1}{T}\mathcal{F}(\nu),
\end{equation}
which effectively reduces Eq.~\eqref{eq:Fgeneral} to a one-dimensional Fourier transform in the difference variable.


We first consider a detector that is static at a fixed radius \( r_0 \) outside the black hole, with fixed angular position \( \Omega_0 \). Its worldline in Schwarzschild-like coordinates can be parameterized as
\begin{equation}
  x^\mu(\tau) = (t(\tau),r_0,\Omega_0),
\end{equation}
where the proper time \( \tau \) and coordinate time \( t \) are related by gravitational redshift,
\begin{equation}
  t(\tau) = \frac{\tau}{\sqrt{f(r_0)}}.
\end{equation}
The static observer is accelerated with respect to free fall and therefore experiences a local temperature related to the Hawking temperature by a Tolman redshift factor~\cite{Birrell:1982ix,Crispino:2007eb}.

Along this trajectory, the coordinate-time difference between two events is
\begin{equation}
  \Delta t = t(\tau) - t(\tau')
           = \frac{\Delta \tau}{\sqrt{f_0}},
  \qquad f_0 := f(r_0),
\end{equation}
where \( \Delta\tau = \tau - \tau' \). Restricting to the QNM sector~\eqref{eq:GQNMsimple}, the Wightman function along the trajectory becomes
\begin{equation}
  G^+_{\rm QNM}(\tau,\tau')
  = \sum_n B_n(r_0)\,
    e^{-i\omega_n \Delta t}
  = \sum_n B_n(r_0)\,
    e^{-i\omega_n \Delta\tau/\sqrt{f_0}}.
  \label{eq:GQNM-static}
\end{equation}

For a long, adiabatic interaction (large effective measurement time), we may approximate the switching function as constant over a large interval and define the QNM contribution to the transition rate as
\begin{equation}
  \dot{\mathcal{F}}_{\rm QNM}(\nu)
  = \int_{-\infty}^{+\infty} d\Delta\tau\,
    e^{-i\nu\Delta\tau}\,
    G^+_{\rm QNM}(\Delta\tau)
  = \sum_n B_n(r_0)
    \int_{-\infty}^{+\infty} d\Delta\tau\,
    e^{-i\nu\Delta\tau}\,
    e^{-i\omega_n \Delta\tau/\sqrt{f_0}}.
  \label{eq:FQNM-start}
\end{equation}
The integral over \(\Delta\tau\) is a Fourier transform of a damped exponential. To make the damping explicit, we write
\begin{equation}
  \omega_n = \Omega_n - i\Gamma_n,\qquad \Gamma_n>0,
\end{equation}
so that
\begin{equation}
  e^{-i\omega_n \Delta\tau/\sqrt{f_0}}
  = \exp\left[-i\frac{\Omega_n}{\sqrt{f_0}}\Delta\tau\right]\,
    \exp\left[-\frac{\Gamma_n}{\sqrt{f_0}}\,\Delta\tau\right].
\end{equation}
For \(\Delta\tau > 0\), the second factor ensures convergence as long as \(\Gamma_n>0\), and a similar argument can be made for \(\Delta\tau < 0\) using the standard \(i\epsilon\)-prescription~\cite{Birrell:1982ix}.

Focusing for the moment on the contribution from \(\Delta\tau>0\), we have
\begin{equation}
  \int_0^{\infty} d\Delta\tau\,
  e^{-i\nu\Delta\tau}
  e^{-i\omega_n \Delta\tau/\sqrt{f_0}}
  = \int_0^{\infty} d\Delta\tau\,
  e^{-\left(\Gamma_n/\sqrt{f_0} + i\,[\nu + \Omega_n/\sqrt{f_0}]\right)\Delta\tau}
  = \frac{1}{\Gamma_n/\sqrt{f_0} + i\bigl(\nu+\Omega_n/\sqrt{f_0}\bigr)}.
  \label{eq:half-line-integral}
\end{equation}
Including the contribution from \(\Delta\tau<0\) and taking the real part yields a Lorentzian in frequency space. It is natural to define the \emph{local} QNM frequency and damping rate as measured by the static observer at \(r_0\):
\begin{equation}
  \omega_n^{\rm (loc)}(r_0)
  := \frac{\Omega_n}{\sqrt{f(r_0)}},
  \qquad
  \gamma_n^{\rm (loc)}(r_0)
  := \frac{\Gamma_n}{\sqrt{f(r_0)}}.
  \label{eq:local-omega-gamma}
\end{equation}
These are simply the global QNM parameters redshifted by the gravitational potential at the detector location.

Up to an overall normalization, we thus find that the contribution of mode \(n\) to the detector rate takes the Lorentzian form
\begin{equation}
  \dot{\mathcal{F}}_n(\nu; r_0)
  \propto \frac{\gamma_n^{\rm (loc)}(r_0)}
   {\bigl(\nu - \omega_n^{\rm (loc)}(r_0)\bigr)^2
    + \bigl(\gamma_n^{\rm (loc)}(r_0)\bigr)^2}.
\end{equation}
Summing over all QNMs, we obtain the QNM-induced contribution to the detector response as
\begin{equation}
  \boxed{
  \dot{\mathcal{F}}_{\rm QNM}(\nu; r_0)
  \simeq \sum_n \mathcal{A}_n(r_0)\,
  \frac{\gamma_n^{\rm (loc)}(r_0)}
   {\bigl(\nu - \omega_n^{\rm (loc)}(r_0)\bigr)^2
    + \bigl(\gamma_n^{\rm (loc)}(r_0)\bigr)^2}
  }
  \label{eq:FQNM}
\end{equation}
where the amplitudes \( \mathcal{A}_n(r_0) \) are proportional to \( |B_n(r_0)|^2 \) and encode the overlap between the QNM spatial profile and the detector location. Equation~\eqref{eq:FQNM} shows that, from the perspective of the static detector, the QNM sector manifests as a set of Lorentzian resonances at the locally redshifted real parts of the QNM frequencies, with linewidths given by the redshifted damping rates.

In a more complete description, the full detector spectrum includes both the QNM sector and the continuum/HBAR contribution. The latter provides an approximately thermal envelope determined by the local Hawking temperature
\begin{equation}
  T_{\rm loc}(r_0) = \frac{T_H}{\sqrt{f(r_0)}}
                    = \frac{\kappa}{2\pi\sqrt{f(r_0)}},
\end{equation}
as emphasized in the HBAR and CQM analyses~\cite{ScullyHBAR1,ScullyHBAR2,Camblong:2022jbg}. In this picture, the QNM resonances appear as sharp peaks superimposed on a broad thermal background, providing a spectrally resolved probe of the black hole's scattering properties.


We briefly comment on the case of a radially infalling detector, which is particularly relevant for the HBAR scenario. Consider, for example, an observer starting from rest at infinity in a Schwarzschild spacetime. The relation between coordinate time \(t\) and proper time \(\tau\) along such a trajectory is more involved, but near the horizon it has the universal leading behavior
\begin{equation}
  t(\tau) \sim -\frac{1}{\kappa}\ln(\tau_H - \tau) + \text{const},
\end{equation}
where \( \tau_H \) is the finite proper time at which the detector crosses the horizon and \( \kappa \) is the surface gravity. Substituting this into the QNM phase factor yields
\begin{equation}
  e^{-i\omega_n t(\tau)}
  \sim (\tau_H-\tau)^{i\omega_n/\kappa},
\end{equation}
so that the QNM contribution to the Wightman function along the trajectory acquires a characteristic power-law dependence on the proper time separation. This is precisely the type of structure that, in the HBAR analysis, leads to a thermal response with temperature \(T_H\), arising from the logarithmic relation between proper time and the outgoing null coordinate~\cite{ScullyHBAR1,ScullyHBAR2}.

Incorporating the QNM phases into the contour integrals that appear in the rigorous evaluation of the response function~\eqref{eq:Fgeneral} for infalling trajectories is technically more involved but conceptually straightforward. The net effect is expected to be a thermal spectrum with additional oscillatory features and modulations governed by the QNM parameters \( \Omega_n \) and \( \Gamma_n \). A full analytic treatment of this case, including the interplay between the QNM-induced resonances and the near-horizon CQM structure, is left for future work.

\section{Single-QNM master equation and lasing threshold}
\label{sec:lasing}

We now move beyond single-detector response and consider a many-body, quantum optical description in which a single dominant QNM is treated as an effective bosonic mode coupled to a cloud of driven two-level atoms. This setup is a natural generalization of the multimode master equations employed in the HBAR program~\cite{ScullyHBAR2,Camblong:2022jbg} and is closely related to the Dicke model of superradiance and lasing~\cite{ScullyZubairy}. Our aim is to derive a Maxwell--Bloch system and a lasing threshold condition in which the QNM damping rate plays the role of cavity loss.


Motivated by the QNM expansion of the Green's function, we model a single dominant QNM by an effective bosonic annihilation operator \(b\) with complex frequency
\begin{equation}
  \omega_Q = \Omega_Q - i\Gamma_Q.
\end{equation}
We work in a frame rotating at frequency \(\Omega_Q\), so that the free Hamiltonian for the QNM mode can be written as
\begin{equation}
  H_{\rm field} = \hbar\Delta_c\,b^\dagger b,
\end{equation}
where \( \Delta_c := \omega_{\rm lab} - \Omega_Q \) denotes a small detuning with respect to some laboratory or driving frequency (we will set \(\Delta_c\approx 0\) near resonance). The imaginary part \(-\Gamma_Q\) corresponds to an effective decay rate, which we incorporate at the level of the master equation through a Lindblad term
\begin{equation}
  \mathcal{L}_{\rm QNM}[\rho]
  = \kappa\,\Big(2b\rho b^\dagger - b^\dagger b\rho - \rho b^\dagger b\Big),
  \qquad \kappa \sim 2\Gamma_Q + \kappa_{\rm extra},
\end{equation}
where \( \rho \) is the density matrix of the combined system and \( \kappa_{\rm extra} \) accounts for additional environmental losses (e.g.\ coupling to other field modes or absorption by external media).

We consider \(N\) identical two-level atoms with transition frequency \( \nu \), described by Pauli operators \( \sigma_j^\pm,\sigma_j^z \) (\( j=1,\dots,N \)) and Hamiltonian
\begin{equation}
  H_{\rm atoms} = \frac{\hbar\nu}{2}\sum_{j=1}^N \sigma_j^z.
\end{equation}
The atom--QNM interaction is modeled by a Dicke-type Hamiltonian
\begin{equation}
  H_{\rm int} = \hbar g\sum_{j=1}^N
    \left(\sigma_j^+ b + \sigma_j^- b^\dagger\right),
\end{equation}
where \( g \) is an effective coupling constant that includes overlap factors between the QNM spatial profile and the atomic cloud, as well as local gravitational redshift factors.

The atoms are taken to interact with additional reservoirs that provide pumping and relaxation processes. At the level of the master equation, these are described phenomenologically by
\begin{equation}
  \mathcal{L}_{\rm atoms}[\rho]
  = \sum_{j=1}^N \left(
      \gamma_\downarrow D[\sigma_j^-]\rho
    + \gamma_\uparrow D[\sigma_j^+]\rho
    \right),
\end{equation}
where
\begin{equation}
  D[L]\rho = L\rho L^\dagger - \frac{1}{2}\{L^\dagger L,\rho\}
\end{equation}
is the standard Lindblad dissipator, and \( \gamma_\uparrow,\gamma_\downarrow \) are excitation and relaxation rates, respectively. A nonzero population inversion corresponds to \( \gamma_\uparrow>\gamma_\downarrow \) and is essential for lasing.

Collecting all contributions, the full master equation for the reduced density matrix of the QNM mode plus atomic cloud reads
\begin{equation}
  \dot{\rho} = -\frac{i}{\hbar}[H_{\rm field}+H_{\rm atoms}+H_{\rm int},\rho]
               + \mathcal{L}_{\rm QNM}[\rho]
               + \mathcal{L}_{\rm atoms}[\rho].
  \label{eq:masterfull}
\end{equation}
This is a standard open quantum system description, now with the key novelty that the cavity mode is identified with a damped QNM of the black hole background.


To derive a lasing threshold condition, we adopt a semiclassical treatment in which we replace operator expectation values by c-numbers and factorize higher-order correlations. We introduce the collective variables
\begin{equation}
  \alpha := \langle b\rangle, \qquad
  P := \sum_{j=1}^N \langle\sigma_j^-\rangle,
  \qquad
  D := \sum_{j=1}^N \langle\sigma_j^z\rangle,
\end{equation}
which play the roles of field amplitude, collective polarization, and total inversion, respectively. Under the usual mean-field approximations (e.g.\ \( \langle b\sigma_j^z\rangle \approx \langle b\rangle\langle\sigma_j^z\rangle \)), the equations of motion derived from Eq.~\eqref{eq:masterfull} reduce to the familiar Maxwell--Bloch equations~\cite{ScullyZubairy}:
\begin{align}
  \dot{\alpha} 
    &= -(\kappa + i\Delta_c)\,\alpha + g P,
    \label{eq:MB1}
    \\
  \dot{P} 
    &= -(\gamma_\perp + i\Delta_a)P + g D \alpha,
    \label{eq:MB2}
    \\
  \dot{D} 
    &= -\gamma_\parallel(D - ND_0)
       - 2g(\alpha^* P + \alpha P^*),
    \label{eq:MB3}
\end{align}
where
\begin{equation}
  \Delta_a := \nu - \Omega_Q
\end{equation}
is the atom--QNM detuning, and \( \gamma_\perp,\gamma_\parallel \) are transverse and longitudinal relaxation rates, respectively, given in terms of \( \gamma_\uparrow,\gamma_\downarrow \) and any additional dephasing processes. The equilibrium inversion per atom in the absence of the field is
\begin{equation}
  D_0 := \frac{\gamma_\uparrow-\gamma_\downarrow}{\gamma_\uparrow+\gamma_\downarrow},
\end{equation}
so that in the absence of coupling to the field we have \(D = N D_0\).

Equations~\eqref{eq:MB1}--\eqref{eq:MB3} describe the coupled dynamics of the QNM mode and the atomic ensemble. The interplay between gain (controlled by \(g^2 N D_0\)) and loss (controlled by \(\kappa\) and \(\gamma_\perp\)) determines whether a coherent macroscopic occupation of the QNM mode can develop.


The ``off'' (non-lasing) state of the system corresponds to
\begin{equation}
  \alpha = 0, \quad P = 0, \quad D = ND_0,
\end{equation}
i.e.\ no coherent field, no polarization, and a uniform inversion. To determine whether this state is stable or unstable to small perturbations, we linearize Eqs.~\eqref{eq:MB1}--\eqref{eq:MB3} around it, neglecting the nonlinear saturation term in Eq.~\eqref{eq:MB3}. We obtain
\begin{align}
  \dot{\alpha} &= -(\kappa + i\Delta_c)\,\alpha + g P,
  \\
  \dot{P}      &= -(\gamma_\perp + i\Delta_a)P + g N D_0 \alpha.
\end{align}
Looking for solutions of the form \( \alpha,P\propto e^{\lambda t} \), we find the eigenvalue equation
\begin{equation}
  \left(
  \begin{array}{cc}
    \lambda + \kappa + i\Delta_c & -g \\
    -g N D_0 & \lambda + \gamma_\perp + i\Delta_a
  \end{array}
  \right)
  \left(
  \begin{array}{c}
    \alpha \\ P
  \end{array}
  \right)
  = 0,
\end{equation}
which leads to the characteristic polynomial
\begin{equation}
  (\lambda + \kappa + i\Delta_c)(\lambda + \gamma_\perp + i\Delta_a)
  - g^2 N D_0 = 0.
  \label{eq:charpoly}
\end{equation}

To obtain a simple threshold condition, we consider the near-resonant case
\begin{equation}
  \Delta_c \approx 0,\qquad \Delta_a \approx 0,
\end{equation}
so that the QNM mode and atomic transition are approximately frequency-matched. Then Eq.~\eqref{eq:charpoly} reduces to
\begin{equation}
  (\lambda + \kappa)(\lambda + \gamma_\perp) - g^2 N D_0 = 0.
\end{equation}
The boundary between stability and instability is reached when the real part of the leading eigenvalue \(\lambda\) crosses zero. Setting \( \lambda = 0 \) yields the lasing threshold condition
\begin{equation}
  \boxed{
  g^2 N D_0^{\rm (thr)} = \kappa\,\gamma_\perp
  }
  \label{eq:D0thr}
\end{equation}
or, equivalently,
\begin{equation}
  D_0^{\rm (thr)} = \frac{\kappa\,\gamma_\perp}{g^2 N}.
\end{equation}
Equation~\eqref{eq:D0thr} is the central result of this section: it directly relates the minimal inversion per atom required for lasing to the cavity loss rate \(\kappa\) and the atomic coherence decay rate \(\gamma_\perp\).

In our black hole context, the effective cavity loss rate is dominated by the QNM damping,
\begin{equation}
  \kappa \sim 2\Gamma_Q + \kappa_{\rm extra},
\end{equation}
where the factor of 2 reflects the relation between the QNM imaginary part and the energy decay rate, and \(\kappa_{\rm extra}\) subsumes additional loss channels. Thus the \emph{imaginary part} of the QNM frequency plays precisely the role of a cavity loss rate that must be compensated by gain from the inverted atomic ensemble. Equation~\eqref{eq:D0thr} therefore furnishes a direct quantum optical interpretation of the QNM damping rate: more strongly damped modes require a larger inversion to reach threshold and lase.

Above threshold, the nonlinear saturation term in Eq.~\eqref{eq:MB3} stabilizes the field amplitude at a finite value, corresponding to a coherent QNM-dominated state. The stationary solutions and their stability properties can be analyzed by solving the steady-state Maxwell--Bloch equations and studying small perturbations around them~\cite{ScullyZubairy}. While such an analysis is straightforward, it is beyond the scope of the present work, where our primary focus is on the threshold condition itself and its dependence on QNM properties.

\section{Schwarzschild case and HBAR--QNM spectroscopy}
\label{sec:Schwarzschild}

We now specialize our framework to the Schwarzschild geometry, where the QNM spectrum is particularly well understood and admits a simple approximation in the eikonal (large-\(\ell\)) limit. This allows us to make our discussion more concrete and to relate the HBAR--QNM features directly to photon-sphere quantities.


For a Schwarzschild black hole of mass \(M\), the metric function is
\begin{equation}
  f(r) = 1 - \frac{2M}{r}.
\end{equation}
The unstable photon sphere is located at the radius
\begin{equation}
  r_c = 3M,
\end{equation}
where null circular geodesics exist. In the eikonal (large-\(\ell\)) limit, the QNM frequencies for various spins are governed by properties of this photon sphere~\cite{Cardoso:2008bp}. Specifically, one finds
\begin{equation}
  \Omega_{n\ell} \approx \ell\,\Omega_c,
  \qquad
  \Gamma_{n\ell} \approx (n+\tfrac{1}{2})\,\lambda_c,
\end{equation}
where
\begin{equation}
  \Omega_c = \sqrt{\frac{f(r_c)}{r_c^2}}
           = \frac{1}{3\sqrt{3}M},
\end{equation}
is the orbital frequency of the null circular orbit, and
\begin{equation}
  \lambda_c = \sqrt{\frac{f(r_c)}{2r_c^2}
   \left(2f(r_c) - r_c^2 f''(r_c)\right)}
  = \frac{1}{3\sqrt{3}M}
\end{equation}
is the corresponding Lyapunov exponent controlling the instability of the orbit. Thus, to leading order in the eikonal expansion, the QNM spectrum takes the simple form
\begin{equation}
  \omega_{n\ell}
  \approx \frac{\ell - i(n+\tfrac{1}{2})}{3\sqrt{3}M}.
\end{equation}
This relation, together with Eq.~\eqref{eq:local-omega-gamma}, provides a direct link between the Schwarzschild QNM parameters and the HBAR--QNM spectroscopic features measured by a detector at radius \(r_0\).

For a static detector at \( r_0 > 2M \), the locally measured QNM frequencies and damping rates are
\begin{equation}
  \omega_{n\ell}^{\rm (loc)}(r_0)
  = \frac{\Omega_{n\ell}}{\sqrt{1-2M/r_0}},
  \qquad
  \gamma_{n\ell}^{\rm (loc)}(r_0)
  = \frac{\Gamma_{n\ell}}{\sqrt{1-2M/r_0}}.
\end{equation}
Substituting into Eq.~\eqref{eq:FQNM}, we see that the detector response exhibits Lorentzian resonances at
\begin{equation}
  \nu \approx \omega_{n\ell}^{\rm (loc)}(r_0),
\end{equation}
with linewidths set by \( \gamma_{n\ell}^{\rm (loc)}(r_0) \). As the detector moves closer to the photon sphere, the amplitudes \(\mathcal{A}_n(r_0)\) are expected to increase, reflecting the stronger overlap with the QNM spatial profiles.

\subsection{HBAR envelope and QNM resonances}

Within the HBAR framework, the detector also experiences a broad thermal-like response governed by the local Hawking temperature. For Schwarzschild, the Hawking temperature is
\begin{equation}
  T_H = \frac{1}{8\pi M},
\end{equation}
and the local Tolman-redshifted temperature at radius \(r_0\) is
\begin{equation}
  T_{\rm loc}(r_0) = \frac{T_H}{\sqrt{1-2M/r_0}}.
\end{equation}
The continuum contribution \( \dot{\mathcal{F}}_{\rm therm}(\nu; T_{\rm loc}) \) arising from near-horizon CQM and the Boulware vacuum structure yields a smooth, approximately Planckian envelope in the detector spectrum~\cite{ScullyHBAR1,ScullyHBAR2,Camblong:2022jbg,Camblong:2005my}. The precise functional form depends on the trajectory and switching, but its qualitative features are those of a thermal distribution at temperature \(T_{\rm loc}\).

Combining the thermal envelope with the QNM-induced resonances, we obtain a schematic form for the total detector response:
\begin{equation}
  \dot{\mathcal{F}}(\nu; r_0)
  \approx \dot{\mathcal{F}}_{\rm therm}(\nu; T_{\rm loc})
        + \dot{\mathcal{F}}_{\rm QNM}(\nu; r_0),
\end{equation}
where \( \dot{\mathcal{F}}_{\rm QNM} \) is given by Eq.~\eqref{eq:FQNM}. We refer to this combination---thermal envelope decorated by discrete QNM Lorentzians---as the \emph{HBAR--QNM spectroscopic fingerprint} of the black hole.

In the single-mode lasing picture developed in Sec.~\ref{sec:lasing}, the Schwarzschild QNM damping rate \( \Gamma_Q \) enters the threshold condition~\eqref{eq:D0thr} via
\begin{equation}
  \kappa \sim 2\Gamma_Q + \kappa_{\rm extra},
\end{equation}
so that the minimal inversion required to stimulate a macroscopic occupation of the QNM mode is directly sensitive to photon-sphere properties. Different black hole spacetimes (e.g.\ regular black holes, non-linear electrodynamics solutions, or modified-gravity backgrounds) generally have different photon-sphere radii, orbital frequencies, and Lyapunov exponents~\cite{Cardoso:2008bp}, and hence different QNM spectra and lasing thresholds. In this way, the combined HBAR--QNM fingerprint offers a robust method of distinguishing between Schwarzschild and other Schwarzschild-like metrics at the level of their microscopic quantum optical signatures.

The distinct physical origins and properties of the thermal and QNM components of the detector signal are summarized in Table~\ref{tab:spectroscopy}.

\begin{table}[h!]
\centering
\caption{Comparison of the HBAR continuum and QNM resonance components of the detector signal. These two components are sourced by distinct physics at different locations: the continuum by near-horizon CQM, and the discrete resonances by photon-sphere dynamics.}
\label{tab:spectroscopy}
\begin{tabular}{|l|l|l|}
\hline
\textbf{Feature} & \textbf{HBAR Continuum Component} & \textbf{QNM Resonance Component} \\
\hline
Signal Type & Thermal continuum & Discrete resonances  \\
Spectral Shape & Broad (Planckian-like)  & Narrow (Lorentzian)  \\
Correlator Origin & $G^+_{\rm cont}(x,x')$ (branch cut)  & $G^+_{\rm QNM}(x,x')$ (poles)  \\
Geometric Origin & Near-horizon region ($r \approx r_h$)  & Photon sphere ($r \approx r_c$)  \\
Governing Physics & Conformal quantum mechanics (CQM)  & Null geodesic instability  \\
Primary Parameters & Hawking temperature $T_H = \kappa / (2\pi)$  & QNM frequencies $\{\Omega_n, \Gamma_n\}$  \\
\hline
\end{tabular}
\end{table}

\section{Figures and qualitative discussion}
\label{sec:figures}

In this section we briefly describe a set of illustrative figures that help visualize the HBAR--QNM framework and its physical implications. These figures can be generated analytically in the eikonal approximation or numerically using standard QNM codes~\cite{Leaver:1986gd,Berti:2009kk}.

\subsection{Schematic geometry and QNM--HBAR setup}

\begin{figure}[t]
  \centering
  \includegraphics[width=0.55\textwidth]{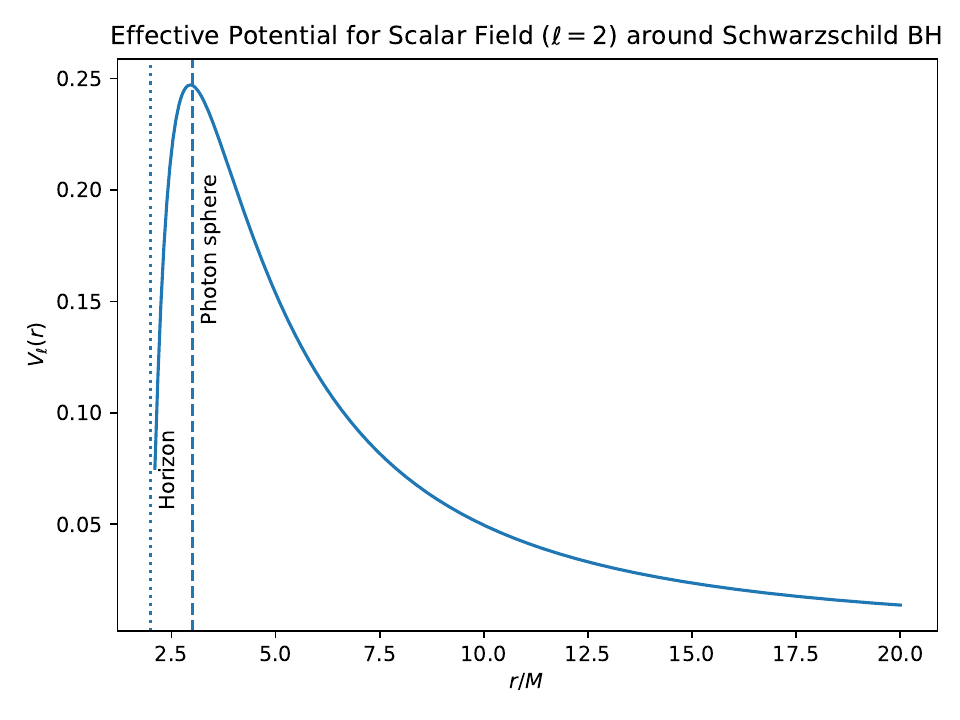}
  \caption{Schematic representation of the setup. A static or infalling cloud of two-level atoms interacts with the scalar field on a black hole background. The effective potential \( V_\ell(r) \) forms a barrier around the photon sphere, supporting quasinormal modes with complex frequencies \( \omega_{n\ell} \). The atoms experience a thermal HBAR background due to near-horizon physics, as well as discrete QNM-induced resonances.}
  \label{fig:schematic}
\end{figure}

Figure~\ref{fig:schematic} shows a black hole with its effective potential barrier around the photon sphere, along with a cloud of two-level atoms. The potential barrier supports quasinormal ringing as waves are transiently trapped and leak out to infinity or into the horizon. At the same time, atoms in the near-horizon region probe the conformal structure responsible for HBAR thermality. This geometric picture highlights the two-scale nature of the problem: the horizon sets the temperature and thermal envelope, while the photon sphere fixes the QNM frequencies and damping that appear as discrete spectral features.

\subsection{Detector spectrum and QNM resonances}

\begin{figure}[t]
  \centering
  \includegraphics[width=0.55\textwidth]{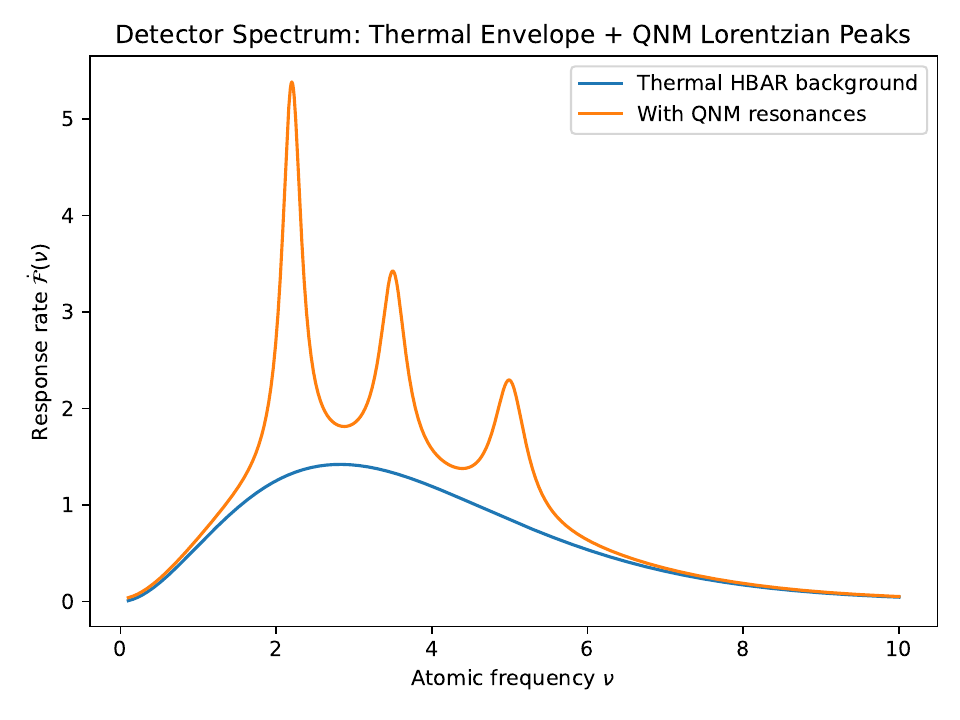}
  \caption{Detector response rate \( \dot{\mathcal{F}}(\nu) \) as a function of atomic frequency \( \nu \) for a static detector at fixed radius \( r_0 \) outside a Schwarzschild black hole. The smooth curve corresponds to the thermal HBAR background at local temperature \( T_{\rm loc} \); the superimposed peaks are Lorentzian resonances at the locally redshifted QNM frequencies \( \omega_{n\ell}^{\rm (loc)}(r_0) \). The widths of the peaks encode the QNM damping rates.}
  \label{fig:spectrum}
\end{figure}

Figure~\ref{fig:spectrum} shows a representative detector spectrum. The broad background is approximately Planckian, with its slope determined by \( T_{\rm loc} \). Superimposed on this envelope are several narrow peaks at discrete frequencies corresponding to \( \omega_{n\ell}^{\rm (loc)}(r_0) \). As the detector moves closer to the horizon, both the thermal envelope and the resonances are redshifted and broadened in a correlated way. Measuring the positions and widths of these peaks as a function of radius provides direct access to \( \Omega_{n\ell}(M) \) and \( \Gamma_{n\ell}(M) \), and hence to the underlying photon-sphere parameters \(\Omega_c\) and \(\lambda_c\).

\section{Conclusions and outlook}
\label{sec:conclusions}

We have proposed and developed a quantum optical framework in which black hole quasinormal modes are probed and manipulated by two-level atoms in the spirit of the HBAR program. Our main results can be summarized as follows: Starting from a QNM expansion of the Wightman function, we derived the QNM contribution to the response of a two-level Unruh--DeWitt detector. For a static detector, the QNM sector yields a series of Lorentzian resonances at the locally redshifted real parts of the QNM frequencies, with widths determined by the redshifted imaginary parts. Treating a single dominant QNM as an effective bosonic mode coupled to an ensemble of pumped two-level atoms, we obtained a Maxwell--Bloch system and derived a lasing threshold condition
  \(
      g^2 N D_0^{\rm (thr)} = \kappa\,\gamma_\perp
  \),
  where the effective cavity loss rate \( \kappa \sim 2\Gamma_Q + \kappa_{\rm extra} \) is directly related to the imaginary part of the QNM frequency. This provides a quantum optical interpretation of QNM damping. Specializing to Schwarzschild, we expressed the leading QNM frequencies and damping rates in terms of photon-sphere quantities and described the resulting HBAR--QNM spectrum: a thermal envelope with superimposed QNM resonances. We argued that this ``HBAR--QNM fingerprint'' can distinguish Schwarzschild from other Schwarzschild-like metrics and encodes both near-horizon CQM and photon-sphere dynamics.

Our analysis opens several avenues for further works. Extending the master equation to multiple QNMs and to entangled atomic ensembles would open an avenue to studying entanglement transfer and decoherence in QNM-dominated regimes, and to exploring connections with non-Hermitian and parity-time symmetric quantum optics, as well as with black-hole laser scenarios in analogue gravity systems. Analogue gravity platforms (e.g.\ optical, acoustic, or Bose--Einstein condensate analogues) and engineered optical cavities with tailored loss profiles might realize aspects of the QNM--lasing scenario explored here, providing experimental access to related phenomena in a controlled setting. Although our framework is theoretical and idealized, the formal parallels between HBAR--QNM spectroscopy and gravitational-wave ringdown suggest that techniques developed here may inspire new ways of interpreting ringdown data, especially in the context of multi-mode reconstruction and tests of the no-hair theorem.

We view the present work as a first step toward a systematic quantum optical theory of black hole ringdown and spectroscopy, in which QNMs are not only passive signatures of spacetime curvature but active participants in a rich non-equilibrium quantum dynamics. Further development along these lines may shed new light on the interplay between gravity, thermodynamics, and quantum information in the strong-field regime.

\begin{acknowledgments}
A. \"O. would like to acknowledge networking support of the COST Action CA21106 - COSMIC WISPers in the Dark Universe: Theory, astrophysics and experiments (CosmicWISPers), the COST Action CA22113 - Fundamental challenges in theoretical physics (THEORY-CHALLENGES), the COST Action CA21136 - Addressing observational tensions in cosmology with systematics and fundamental physics (CosmoVerse), the COST Action CA23130 - Bridging high and low energies in search of quantum gravity (BridgeQG), and the COST Action CA23115 - Relativistic Quantum Information (RQI) funded by COST (European Cooperation in Science and Technology). A. \"O. also thanks to EMU, TUBITAK, ULAKBIM (Turkiye) and SCOAP3 (Switzerland) for their support.
\end{acknowledgments}

\bibliography{ref}

\end{document}